\documentclass[prd,superscriptaddress,amsfonts,amssymb,amsmath,showpacs]{revtex4-2}
\usepackage[utf8]{inputenc}
\usepackage{siunitx}
\usepackage[version=4]{mhchem}
\usepackage{url}
\usepackage{tensor}
\usepackage{longtable}
\usepackage{float}
\usepackage{mathtools}
\usepackage{amssymb}
\usepackage{amsmath}
\usepackage{color, colortbl}
\usepackage{adjustbox}
\usepackage{tikz}
\usepackage[pages=some]{background}
\usepackage{hyperref}
\usepackage{accents}
\usepackage{todonotes}
\usepackage{eqnarray}
\usepackage{makecell}
\usepackage[margin=1in]{geometry}
\usepackage{mathrsfs}
\usepackage{ulem}

\hypersetup{
    colorlinks=true,
    linkcolor=blue,
    filecolor=magenta,
    citecolor=red
}
\usepackage{comment}

\newcolumntype{C}{>{\centering\arraybackslash}X}
\newcolumntype{x}[1]{>{\centering\arraybackslash\hspace{0pt}}p{#1}}

\newcommand{\udt}[3]{#1^{#2}_{\phantom{#2}#3}}
\newcommand{\udut}[4]{#1^{#2\phantom{#3}#4}_{\phantom{#2}#3\phantom{#4}}}

\newcommand{\dut}[3]{#1_{#2}^{\phantom{#2}#3}}
\newcommand{\dudt}[4]{#1_{#2\phantom{#3}#4}^{\phantom{#2}#3}}

\newcommand{\lc}[1]{\accentset{\circ}{#1}}
\newcommand{\dd}{\text{d}}

\begin{document}

\title{White Dwarf Envelops and Temperature Corrections in Exponential \texorpdfstring{$f(T)$}{} Gravity}

\author{Gabriel Farrugia}
\email{gfarr02@um.edu.mt}
\affiliation{Institute of Space Sciences and Astronomy, University of Malta, Msida, Malta}
\affiliation{Department of Physics, University of Malta, Msida, Malta}

\author{Carlos Gafa'\footnote{Corresponding author}}
\email{carlos.gafa.19@um.edu.mt}
\affiliation{Institute of Space Sciences and Astronomy, University of Malta, Msida, Malta}
\affiliation{Department of Physics, University of Malta, Msida, Malta}

\author{Jackson Levi Said}
\email{jackson.said@um.edu.mt}
\affiliation{Institute of Space Sciences and Astronomy, University of Malta, Msida, Malta}
\affiliation{Department of Physics, University of Malta, Msida, Malta}

\begin{abstract}
Compact stars have long served as a test bed of gravitational models and their coupling with stellar matter. In this work, we explore the behavior of an exponential model in $f(T)$ gravity through the Tolman–Oppenheimer–Volkoff equation. This is performed for different envelope thicknesses. Finally, constraints on the models parameters are obtained, which are comparable to the results obtained using cosmological survey data. This consistency across the strong astrophysical and weak cosmological scales shows reasonable viability of the underlying model.
\end{abstract}

\maketitle

\section{Introduction}\label{sec:intro}
General relativity (GR) offers a robust framework in which to study compact stars and to constrain their properties using observational measurements. These stars are the result of gravitational collapse that becomes dominated by GR effects. However, GR is known to express a UV completeness problem \cite{AlvesBatista:2023wqm,Addazi:2021xuf} which is expected to play an important role in such stellar systems. Moreover, as part of the cosmological concordance model, it has several issues that merit an assessment of alternatives which range from foundational issues related to divergences in the value of the cosmological constant within GR \cite{Weinberg:1988cp}, as well as the growing challenge of cosmological tensions between survey data \cite{Abdalla:2022yfr}. Given the plethora of concerns related to the behavior of GR in general, and its strong field limit in particular, compact stars offer an ideal testbed in which to constrain possible deviations from this baseline model. There has been a diversity of responses to the growing theoretical and observational challenges to the concordance model and its underlying GR description of gravity including developing gravitational models beyond GR \cite{Addazi:2021xuf,CANTATA:2021ktz,Cai:2019bdh,LeviSaid:2021yat}. In light of this, there has been a revival in interest in many theories beyond GR with renewed interest in their astrophysical consequences.

Teleparallel gravity (TG) provides an interesting geometric framework in which the curvature associated with the Levi-Civita connection is exchanged with the torsion produced by the teleparallel connection \cite{Bahamonde:2021gfp,Krssak:2018ywd,Cai:2015emx}. Both connections are metric compatible with the main outcome being a new framework on which to construct physical theories beyond GR. On the other hand, TG can produce a realization in which it is dynamically equivalent to GR, namely the teleparallel equivalent of general relativity (TEGR) \cite{Aldrovandi:2013wha}. At the level of the classical equations of motion, this is identical to GR while it may differ in the non-classical regime \cite{Mylova:2022ljr}. 

Analogous to curvature-based theories of gravity \cite{CANTATA:2021ktz}, TG has produced a plethora of theories beyond TEGR \cite{Bahamonde:2021gfp,Krssak:2018ywd,Cai:2015emx}. The most straightforward generalization of TEGR is that of $f(T)$ gravity where the Lagrangian is an arbitrary function of the TEGR Lagrangian \cite{Ferraro:2006jd,Ferraro:2008ey,Bengochea:2008gz,Linder:2010py,Chen:2010va,Bahamonde:2019zea,Paliathanasis:2017htk,Farrugia:2020fcu,Bahamonde:2021srr,Bahamonde:2020bbc,Bahamonde:2022ohm}. Here, the scalar $T$ is called the torsion scalar, and plays the role of TEGR Lagrangian. Along a similar vain as $f(\lc{R})$ gravity, $f(T)$ gravity offers interesting directions in which to confront observational and theoretical challenges with concordance cosmology. On the other hand, unlike $f(\lc{R})$ gravity, $f(T)$ gravity is not bound by the Lovelock theorem and propagates second order field equations which is intriguing for a variety of reasons including solvability and numerical stability. There are a variety of other directions in which to consider cosmologies beyond TEGR such as other generalizations \cite{Bahamonde:2015zma,Kofinas:2014aka}, scalar-tensor frameworks \cite{Bahamonde:2022lvh,Bahamonde:2019shr,Bahamonde:2019ipm,Bahamonde:2020cfv}, and others, but $f(T)$ gravity provides a good testing ground for toy models.

Compact stars have a long history is being explored in TG since they offer a strong field analogue to the weak field cosmological work for individual models in this framework \cite{Bahamonde:2021gfp,Capozziello:2011et,Olmo:2019flu}. One of the first robust works on the topic was Ref.~\cite{Boehmer:2011gw} where relativistic stars more general were studied in $f(T)$ gravity with the aims of showing that they firstly exist in this gravitational framework together with some exploration of the structure through their expression of energy conservation. The topic was more exhaustively explored for the particular case of a quadratic model in Ref.~\cite{Ilijic:2018ulf} where a polytropic equation of state was examined which led to the evolution plots expressing the exterior geometry of these types of stars in detail. In Refs.~\cite{Ilijic:2020vzu,Horvat:2014xwa,Horvat:2015qva} this approach was extended to Boson stars which offer an intriguing avenue for understanding more extreme compact stars. On the other hand, neutron stars have been studied in Refs.~\cite{Lin:2021ijx,Pace:2017aon,Pace:2017dpu}. Since this time, a lot of effort has been put into the problem of further developing the underlying physical framework as well as the numerical techniques to explore compact stars in $f(T)$ gravity \cite{deAraujo:2021pni,Fortes:2021ibz,delaCruz-Dombriz:2014zaa,Velay-Vitow:2017odc} with a focus on more realistic stars in this regime \cite{Nashed:2020kjh,deAraujo:2022spy,deAraujo:2023tzj}.

On this last point, we are motivated to produce more realistic models of compact stars in $f(T)$ gravity with an interest in also fitting the model parameters using observational data. It is crucial to understand the strong field effects of any modification to GR since this growing field of study may yield more testable predictions in contrast to the concordance model. To achieve this, we first motivate TG and its $f(T)$ gravity extension, introduced in Sec.~\ref{sec:f_T_background}. The master equations of our compact star scenarios are explored in Sec.~\ref{sec:comp_star_derivations} where the Tolman–Oppenheimer–Volkoff (TOV) equation is derived and then applied for a white dwarf background. Observational data used to fit the model parameters are discussed in Sec.~\ref{sec:data}, followed by the respective constraint analysis in Sec.~\ref{sec:constraint_analysis}. The conclusion where we summarize our main results and give an outlook for future work is provided in Sec.~\ref{sec:conclu}. 

\section{\texorpdfstring{Overview of $f(T)$}{} Gravity} \label{sec:f_T_background}

TG bases on describing gravitation as a manifestation of spacetime torsion as opposed to curvature as described in GR. To achieve this, teleparallel based theories replace the Levi-Civita connection, $\mathring{\Gamma}^{\sigma}_{\mu\nu}$\footnote{From here onwards, overdots shall denote quantities calculated with the Levi-Civita connection.} with the torsionful, curvatureless  Weitzenb\"{o}ck connection, $\Gamma^{\sigma}_{\mu\nu}$, which also satisfies the metricity condition \cite{Krssak:2018ywd,Bahamonde:2021gfp}. 

The tetrad, $\udt{e}{A}{\mu}$, becomes the fundamental dynamical variable of the theory, replacing the metric tensor in standard curvature based theories. Their role is to map between the global manifold spacetime geometry (which coordinates are denoted by Greek indices) with the local (tangent) spacetime geometry endowed by the Minkowski spacetime (which coordinates are denoted by Latin indices). The tetrads satisfy the orthogonality relations 
\begin{align}
    \udt{e}{A}{\mu}\dut{E}{B}{\mu}=\delta^B_A\,,& &\udt{e}{A}{\mu}\dut{E}{A}{\nu}=\delta^{\nu}_{\mu}\,,
\end{align}
where $\dut{E}{A}{\mu}$ represents the inverse tetrad. Consequently, a link between global and local geometries can be established by virtue of the metric relations
\begin{align}\label{eq:metric_tetrad_eq}
    g_{\mu\nu}=\udt{e}{A}{\mu}\udt{e}{B}{\nu}\eta_{ab}\,,& &\eta_{ab}=\dut{E}{A}{\mu}\dut{E}{B}{\nu}g_{\mu\nu}\,,
\end{align}

The Weitzenb\"{o}ck connection is defined as \cite{Weitzenbock:1923efa}
\begin{equation}
    \Gamma^{\sigma}_{\mu\nu} := \dut{E}{A}{\sigma}\partial_{\mu}\udt{e}{A}{\nu} + \dut{E}{A}{\sigma}\udt{\omega}{A}{B\mu}\udt{e}{B}{\nu}\,,
\end{equation}
where $\udt{\omega}{a}{b\mu}$ represents the spin connection which has connotations with regards to inertial effects. Observe that for a given metric, by Eq.~\eqref{eq:metric_tetrad_eq}, the tetrad generating that metric is not unique. To maintain the respective TG theory covariant, the spin connection must be appropriately chosen to counteract the choice of tetrad. Additionally, the clear incorporation of the Minkowskian geometry results in potential spurious inertial effects. Indeed, from Eq.~\eqref{eq:metric_tetrad_eq}, if the global spacetime is Minkowski spacetime, the relation would be satisfied provided the tetrads are precisely the local Lorentz transformations (LLT). To properly account for these inertial contributions, the spin connection is chosen to be the so called inertial spin connection which is flat and  satisfies~\cite{Bahamonde:2021gfp}
\begin{align}\label{eq:flat_spin_connection}
    \partial_{[\mu}\udt{\omega}{A}{|B|\nu]} + \udt{\omega}{A}{C[\mu} \, \udt{\omega}{C}{|B|\nu]} = 0\,,
\end{align}
where square brackets denote the antisymmetric operator. To maintain consistency for such flat spacetimes, the inertial spin connection must be
\begin{align}\label{eq:spin_connection}
    \udt{\omega}{A}{B\mu} = \udt{\Lambda}{A}{C}\,\partial_{\mu}\dut{\Lambda}{B}{C}\,.
\end{align}
Here $\udt{\Lambda}{A}{B}$ denotes Lorentz boosts and rotations~\cite{Aldrovandi:2013wha}. Consequently, one can always choose a Lorentz frame which spin connection is zero. Such tetrads are called proper tetrads \cite{Aldrovandi:2004db} and the theory is said to be in the Weitzenb\"{o}ck gauge. In what follows, this will be assumed \textit{a priori} as the chosen tetrads shall satisfy such gauge. Thus, the resulting field equations shall omit contributions arising from the spin connection.

As TG theories require a torsional measure of spacetime, a tensor analogous to the Riemann tensor in GR (which is used as a measure of curvature) is required. Since the Riemann tensor vanishes under the Weitzenb\"{o}ck connection, the torsion tensor is instead introduced~\cite{Aldrovandi:2004db,Krssak:2018ywd,Bahamonde:2021gfp} 
\begin{align} \label{eq:torsion_tensor}
    \udt{T}{A}{\mu\nu} = \Gamma^{A}_{\nu\mu} - \Gamma^{A}_{\mu\nu}\,,
\end{align}
which represents the field strength of gravitation (as a consequence of the gauge formulation of the tetrad field) \cite{Aldrovandi:2013wha}, and it transforms covariantly under both diffeomorphisms and LLTs. In order to define a suitable gravitational Lagrangian for TG, the torsion tensor is used to construct a suitable scalar analogous to the Ricci scalar, $R$. While various torsional scalar components can be obtained from contractions of the torsion scalar (such as New General Relativity \cite{Hayashi:1979qx} - see \cite{Bahamonde:2021gfp} for further details on the topic), one particular choice is able to provide the same underlying dynamics of GR, i.e. TEGR. Such scalar, referred to as the torsion scalar, $T$, is defined as
\begin{equation}
    T = \frac{1}{4} T^{\lambda\mu\nu}T_{\lambda\mu\nu} + \frac{1}{2}T^{\lambda\mu\nu}T_{\mu\lambda\nu} - \dut{T}{\lambda\mu}{\lambda} \udt{T}{\nu\mu}{\nu}\,.
\end{equation}
which is related to the Ricci scalar (equipped with the Levi-Civita connection) via \cite{Hayashi:1979qx,Hehl:1976kj}
\begin{equation}\label{Ricci_torsion_equiv}
    \mathring{R} = -T + \frac{2}{e}\partial_{\mu}\left(e\udut{T}{\sigma}{\sigma}{\mu}\right) = -T + 2\mathring{\nabla}_{\mu}\left(\udut{T}{\sigma}{\sigma}{\mu}\right)\,,
\end{equation}
where $e=\text{det}\left(\udt{e}{a}{\mu}\right)=\sqrt{-g}$, and $B:=2\mathring{\nabla}_{\mu}\left(\udut{T}{\sigma}{\sigma}{\mu}\right)$ is a boundary term. This total divergence term ensures the equivalence between the equations of motion between GR and TEGR. It is also remarked that the torsion scalar can be rewritten in the form
\begin{equation}\label{torsion_scalar_def}
    T = \dut{S}{A}{\mu\nu}\udt{T}{A}{\mu\nu}\,,
\end{equation}
where 
\begin{align}
    \udt{K}{\sigma}{\mu\nu} &:= \Gamma^{\sigma}_{\mu\nu} - \mathring{\Gamma}^{\sigma}_{\mu\nu} = \frac{1}{2}\left(\dudt{T}{\mu}{\sigma}{\nu} + \dudt{T}{\nu}{\sigma}{\mu} - \udt{T}{\sigma}{\mu\nu}\right)\,, \\
    \dut{S}{A}{\mu\nu} &:= \frac{1}{2}\left(\udt{K}{\mu\nu}{A} - \dut{E}{A}{\nu}\udt{T}{\alpha\mu}{\alpha} + \dut{E}{A}{\mu}\udt{T}{\alpha\nu}{\alpha}\right)\,,
\end{align}
representing the contorsion and superpotential tensors respectively. The latter has been shown to have a possible relation to describing the energy-momentum tensor for gravitation \cite{Aldrovandi:2004db,Koivisto:2019ggr}. 

Consequently, TEGR is obtained via the action
\begin{equation}
    \mathcal{S}_{\text{TEGR}} = -\frac{1}{2\kappa^2}\int \mathrm{d}^4 x\, eT + \int \mathrm{d}^4 x\, e\, \mathcal{L}_m\,,
\end{equation}
where $\kappa^2=8\pi G$ and $\mathcal{L}_m$ is the matter Lagrangian. While the TEGR Lagrangian yields the same underlying dynamics as GR, the theories are still fundamentally different. Nonetheless, to address the observational issues in GR, one possibility is to generalise the gravitational action, such as $f(\mathring{R})$ gravity. In a similar vein, one can generalise TEGR to introduce an arbitrary function of the torsion scalar, hence leading to $f(T)$ gravity. These theories are fundamentally distinct.

Firstly, contributions from the torsion scalar lead to second-order field equations, in contrast to generalised Ricci scalar actions, namely $f(\mathring{R})$ gravity, which result in fourth-order equations. This occurs due to a weakening of Lovelock's theorem in TG \cite{Gonzalez:2015sha,Bahamonde:2019shr,Lovelock:1971yv}. Order can be recovered when introducing the boundary term as it embodies this fourth-order nature. Indeed, one can recover the $f(\mathring{R})$ gravity field equations as a sub-case of $f(T,B)$ gravity which incorporates a generalisation of both torsion scalar and boundary term in the gravitational action. Moreover, $f(T)$ gravity shares other properties with GR including having the same gravitational wave polarisations  \cite{Farrugia:2018gyz,Capozziello:2018qcp}, and being Gauss-Ostrogradsky ghost free \cite{Krssak:2018ywd}.

Hence, the $f(T)$ gravity action is formally defined to be
\begin{equation}\label{f_T_action}
    \mathcal{S}_{f(T)} = \frac{1}{2\kappa^2}\int \mathrm{d}^4 x\, e\left(-T+f(T)\right) + \int \mathrm{d}^4 x\, e \,\mathcal{L}_m\,,
\end{equation}
which, following taking variations with the tetrad results in the following field equations \cite{Bahamonde:2015zma,Farrugia:2018gyz}
\begin{align}
    2\dut{S}{A}{\mu\lambda}\partial_{\mu}f_T - \frac{2}{e} \left(1 - f_T\right) \partial_{\mu}\left(e\dut{S}{A}{\mu\lambda}\right) + 2 \left(1- f_T\right) \udt{T}{\sigma}{\mu A}\dut{S}{\sigma}{\lambda\mu} - \frac{1}{2}\left(-T + f\right)\dut{E}{A}{\lambda} = \kappa^2 \dut{\Theta}{A}{\lambda}\,,
\end{align}
where subscripts denote derivatives, and $\Theta_{\nu\lambda} = \udt{e}{A}{\nu}\dut{\Theta}{A}{\lambda}$ is the regular energy-momentum tensor for matter.

\section{Compact stars in \texorpdfstring{$f(T)$}{} Gravity} \label{sec:comp_star_derivations}

The governing equations for compact stars are developed in this section, together with a deeper look into the case of white dwarf stars and the evolution of their mass-radius for the case of an exponential correction term in the $f(T)$ Lagrangian model.

\subsection{TOV Equations}\label{sec:TOV_eqns}

The TOV equations represent the system of equations associated with compact star systems. In our case, we assume a spherically symmetric stellar system. We begin by considering the static spherically symmetric metric 
\begin{equation}\label{met}
	\dd s^2=e^{2\phi(r)}\dd t^2-e^ {2\Lambda(r)} \dd r^2-r^2\left(\dd\theta^2+\sin^2\theta \, \dd\psi^2\right) \,.
\end{equation}
Moreover, we assume that the energy-momentum tensor of the star is given by that of a perfect isotropic fluid with energy density $\epsilon$ and pressure $P$. Then, the equations of motion in $f(T)$ gravity for such a metric are \cite{Bahamonde:2021gfp}
\begin{align}\label{eom0}
	\kappa \epsilon &= \dfrac{e^{-2\Lambda}\left[-4 f_T \left(e^\Lambda-1+\Lambda' r + \phi' \left(e^\Lambda-1\right)r\right)-r\left(4 f_T'\left(e^\lambda-1\right)+e^{2 \Lambda}f r\right)\right]}{r^2}\,,\\\label{eom1}
	-\kappa P &=-\frac{f}{2}-\frac{2 f_T e^{-2\Lambda}\left(e^\Lambda\left(1+\phi' r\right)-1-2 \phi' r\right)}{r^2}\,,	\\
	-\kappa P &=\frac{e^{-2\Lambda}\left[r\left(2 f_T'\left[1-e^\Lambda+\phi' r\right]\right)\right]}{2r^2}\nonumber\\
	&+\frac{2 f_T\left[1+e^{2\Lambda}-2e^\Lambda\left(1+\phi' r\right)+r\left(3\phi'-\Lambda'+\left[\phi''+\phi'\left(\phi'-\Lambda'\right)\right]r\right)\right]}{2r^2}\label{eom2} \, ,
\end{align}
where $'$ denotes differentiation with respect to radial coordinate $r$, and the torsion scalar is given by
\begin{equation}\label{torsionscalar}
	T=-\frac{2 e^{-2 \Lambda}\left(e^\Lambda-1\right)\left(1+ 2 \phi' r-e^{\Lambda}\right)}{r^2}.
\end{equation}
In order to solve the TOV equations to obtain the mass-radius relationship for a given white dwarf for an exponential $f(T)$ model, the following procedure is followed. First, the torsion scalar and its derivatives, as well as the relevant $f(T)$ and its derivatives are substituted into the field equations to obtain a set of equations in terms of $\phi$ and $\Lambda$, and their spatial derivatives. Subsequently, Eqs.~\eqref{eom1} and \eqref{eom2} are solved for $\phi''$ and $\Lambda'$. These two differential equations are then coupled with the equation for conservation of momentum,
\begin{equation}\label{conserv}
	\frac{\dd P}{\dd r}= -\phi'\left(P+\epsilon\right) \, , 
\end{equation}
which replaces the use of Eq.~\eqref{eom0} with a simpler equation. The resulting system of differential equations, when coupled to an appropriate equation of state, are then be solved for $P$, $\phi'$ and $\Lambda$ using standard numerical methods, with boundary conditions 
\begin{align}\label{eq:boundary_cond}
    \phi'(0) = 0 &\, ,\quad&\Lambda(0)=0 &\,,\quad&   \quad& P(R)=0\,, & P(0)=P_0    \, ,
\end{align}
where $R$ is the radius of the stellar configuration \cite{Ilijic:2018ulf}. The first two conditions set the gravitational force to zero at center of symmetry, the third marks the outer boundary of the star as the point where the the pressure goes to zero, and the last condition is the pressure of at the center which will be used as a parameter to generate the mass-radius relationship.

This approach to solving the TOV equations differs from that described in \cite{Ilijic:2018ulf}, in that it makes use of the conservation of momentum equation. Such an improvement was first implemented in \cite{deAraujo:2022spy}, and was reported to improve numerical stability, especially at high central densities. Moreover, this improves upon the route taken in \cite{deAraujo:2022spy} as it does not require one to obtain an algebraic relationship for $\phi$ through the equations of motion. This allows for greater flexibility in the choice of $f(T)$ in terms of numerical solvability. Particularly, we will demonstrate this through the model $f(T)=-T-a (1-e^{-b T})$, which is known as the variant Linder model. This has shown promising signs when constrained against cosmological data both at background level \cite{Benetti:2020hxp,Xu:2018npu,Wang:2020zfv,Briffa:2021nxg,Briffa:2023ern}, as well as in the perturbative sector \cite{Nesseris:2013jea,Anagnostopoulos:2019miu,Briffa:2023ozo}. Moreover, the model has some attractive numerical features in the context of relativistic stars.

As noted by \cite{Olmo:2019flu,Ilijic:2018ulf}, only the particle number, and by extension the rest mass, is known unambiguously in $f(T)$ gravity, which do not correspond to any observable quantity. In fact other definitions for the mass in $f(T)$ gravity have cropped up in the literature \cite{mass}. As there is no definitive conclusion on the mass in $f(T)$ gravity, we use the standard GR mass,
\begin{equation}
    \frac{\dd M}{\dd r}=4\pi \epsilon r^2 \,,
\end{equation}
to be able to compare to experimental data.

\subsection{White Dwarf Equation of State}

In this work, we will consider white dwarfs to test the $f(T)$ model. This will be done by solving the TOV equations for the Chandrasekhar white dwarf equation of state (EoS), where the pressure $P$ and energy density $\epsilon$ are given by \cite{sphap}
\begin{align}
    P_{\text{CH}}(x)&=A \sqrt{1+x^2}\left(2x^3-3x\right) +3\sinh^{-1}(x)\,, \\
    \epsilon_{\text{CH}}&= A\left[\frac{8}{q}x^3+\left(\sqrt{1+x^2}\left(6x^3+3x\right)-3\sinh^{-1}(x)\right)\right] \,,
\end{align}
where $x_F=\frac{p_F}{m_e c}$ is the dimensionless Fermi momentum, $A=\frac{\pi {m_e}^4c^5}{3 h^3}$, $q=\frac{m_e}{\mu_e m_p}$, and $\mu_e$ is the number of nucleons per electron, where the speed of light $c$ was included for completeness. Such an EoS assumes that the white dwarf consists of a uniform sea of non-interacting, zero-temperature, relativistic electrons, free from the non-interacting nuclear background whose degeneracy pressure supports itself against gravity. Notably, this EoS depends on the composition only through $\mu_e$ which is 2 for all stars under consideration here: carbon, oxygen, helium. 

Nonetheless, while the Chandrasekhar model covers the basic physical principle through which white dwarfs support themselves, further refinements are required if we are to accurately model white dwarf stars. A more realistic zero-temperature white dwarf model was developed by Hamada and Salpeter \cite{hamada,salpeter}. In this model, the classical Coulomb effects are taken into account, the Thomas-Fermi correction, and the exchange between energy and the correlation energy are considered. However, this EoS depends on the atomic number of core under consideration, therefore breaking the degeneracy between oxygen, carbon and helium cores. 

Even when Salpeter's corrections \cite{Koester_1990} are taken into account, the EoS still describes white dwarf stars at zero temperature which is not the case for realistic stars. Another characteristic not taken into account through Salpeter's EoS is the effect of the envelop on the radius of a white dwarf. This envelop consists of non-degenerate matter, usually hydrogen or helium, and therefore cannot be modeled by zero-temperature models such as Salpeter's EoS. The effects that the white dwarf temperature and envelop size have on it radius, for a given mass, are found using evolutionary models, such as those by Wood \cite{wood} and Fontaine \cite{fonta}.

To model these effects we will take a similar approach to Ref.~\cite{wdpaper}. We assume that the change in the radius of a white dwarf, $R_{\text{Evolutionary Model}}$, due to temperature, $\theta$, and envelope effects, from that predicted by Salpeter's EoS, $R_{\text{HS}}$, is independent of the underlying theory of gravity, such that this deviation is given by
\begin{equation}
    \Delta R(\theta,M)\approx R_{\text{HS}}(\theta=0,M,f(T)=-T) - R_{\text{Evolutionary Model}}(\theta,M)
\end{equation}
This will then be taken into account in our data analysis by shifting the radii in our data through 
\begin{equation}\label{correction}
    R_{\text{Shifted}}(\theta,M) = R_{\text{Data}}+\Delta R(\theta,M) \,.
\end{equation}
where the difference is calculated using evolutionary models from Ref.~\cite{Panei:1999ji}. These evolutionary models where chosen because of their wide mass and temperature ranges. The mass-radius relationships in these models assume either no envelop or a hydrogen envelop of $M_{\ce{H}}/M = 3 \times 10^{-4}$ for helium core stars, and either a thin hydrogen envelop of $M_{\ce{H}}/M = 1 \times 10^{-5}$  or a thick helium envelop of $M_{\ce{He}}/M = 1 \times 10^{-2}$ for carbon/oxygen stars.

\subsection{Behaviour of White Dwarf Mass-Radius Relationships in $f(T)$ Gravity}

Mass-radius relationships have been produced in $f(T)$ gravity with quadratic and cubic modifications in \cite{novelapp,deAraujo:2023tzj}. Using the method described in Section~\ref{sec:TOV_eqns}, we will extend these results to an exponential modification, the variant Linder model $f(T)=-T-a (1-e^{ b T})$, where $a$ and $b$ are model parameter constants. This model, motivated by exponential models in $f(R)$ gravity, corresponds to the model presented in \cite{Nesseris:2013jea} with the constant $T_0$ being absorbed as part of the parameter $b$. Moreover, for an initial qualitative analysis of the model's behaviour, we will make use of the Chandrasekhar EoS.

\begin{figure}[H]
	\caption{The white dwarf mass-radius relationship in $f(T)=-T-a (1-e^{ b T})$ gravity. The model parameters $a$ and $b$ are taken arbitrarily such that the different behaviours are shown. Here $R_o=\sqrt{\frac{1}{\kappa A}}$.}
	\centering
	\label{compall}
	\includegraphics[angle=0 ,width=0.75\textwidth,keepaspectratio]{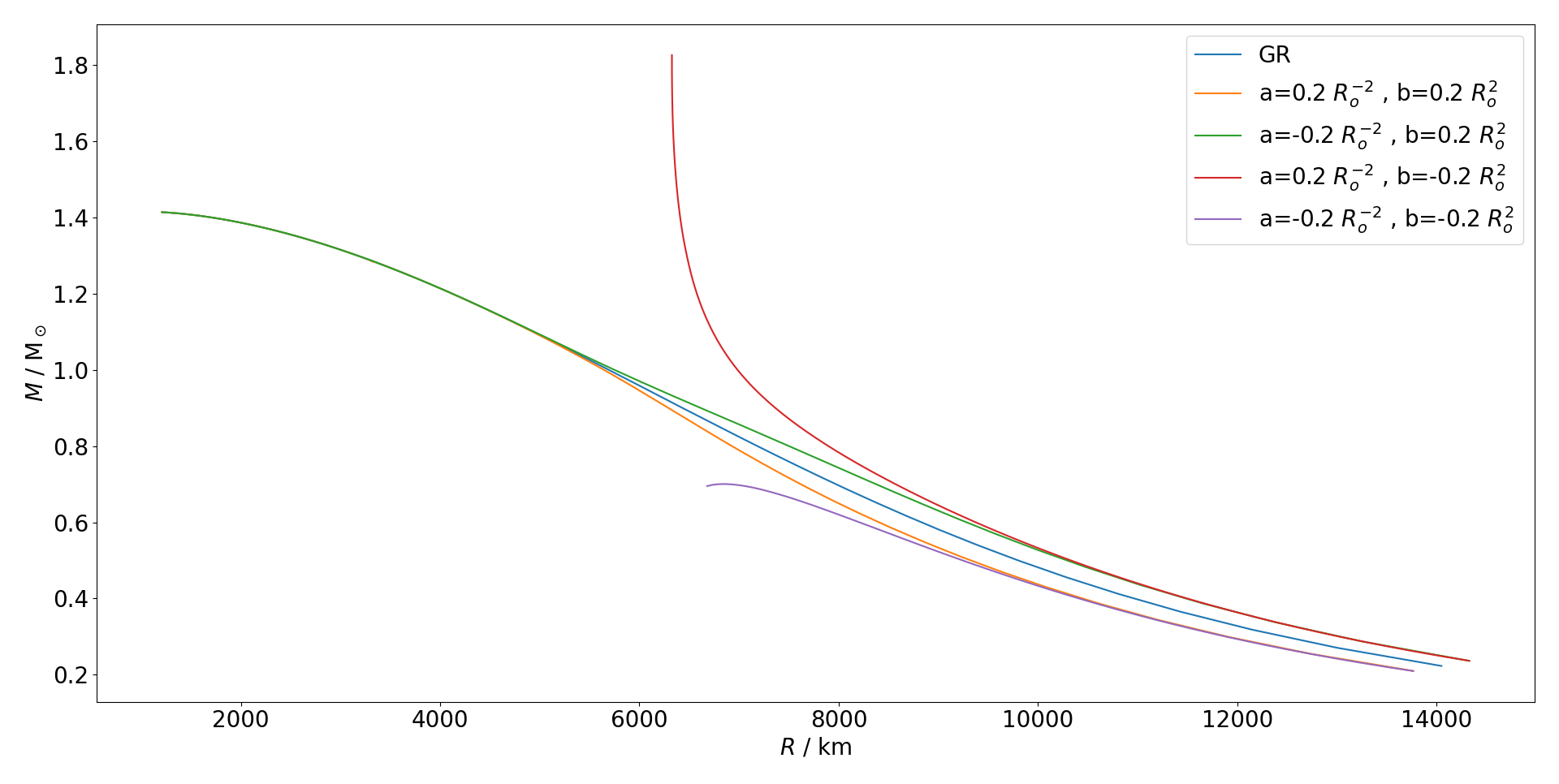}
\end{figure}

From Fig.~\ref{compall}, it is noted that for positive values of $b$ the mass-radius relationship tends to that obtained from GR as the central density increases. To see why this is so, we first note that the torsion scalar inside the star is negative for this metric signature, and so the exponential term becomes negligible at large central densities. This is not the case for when $b$ is negative and we observe different behaviours at large central densities from their counterparts with positive $b$.

\begin{figure}[H]
	\caption{Torsion scalar along the radius $r$ of a white dwarf with a central dimensionless Fermi momentum of $x_0=0.5$, that is corresponding to the first point in the mass-radius relationship presented in Fig. \ref{compall}.}
	\centering
	\label{tor05}
	\includegraphics[angle=0 ,width=0.75\textwidth,keepaspectratio]{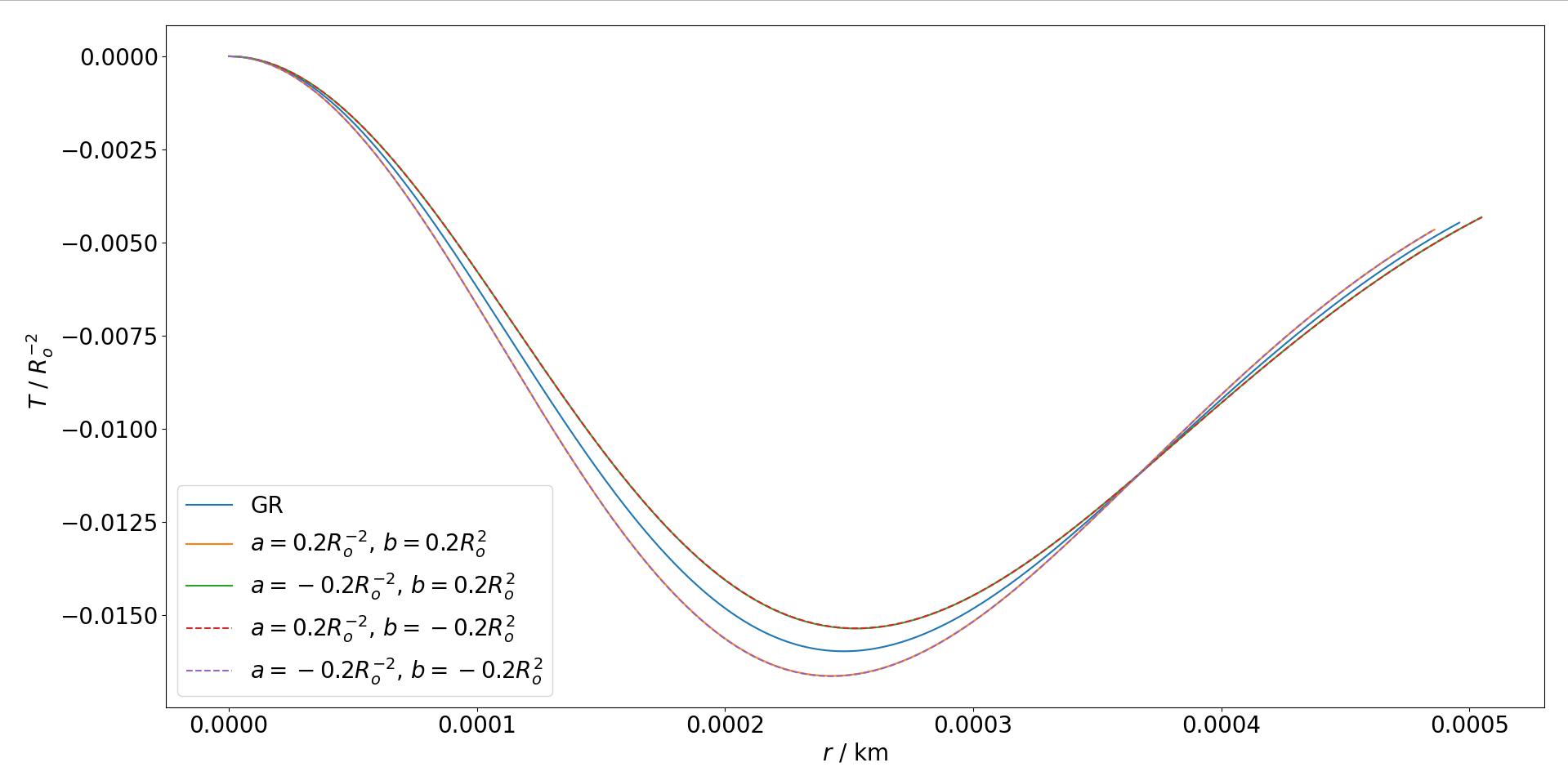}
\end{figure}

These differences in behaviour may also be observed through the torsion scalar. In Fig.~\ref{tor05}, we plot the torsion scalar inside white dwarfs with a low central density. Here it is observed that the models that lie below the GR mass-radius relationship have a smaller peak absolute value of the torsion scalar along the radius of the star, indicating weaker gravity. The opposite holds for the models whose mass-radius relationship lies above the GR one. 

\begin{figure}[H]
	\caption{Torsion scalar along the radius $r$ of a white dwarf with a central dimensionless Fermi momentum of $x_0=1.8$, corresponding to the turning point in the mass-radius curve with parameter values of $a=-0.2 R_o^{-2}$ and $b=-0.2 R_o^{2}$.} 
	\centering
	\label{tor18}
	\includegraphics[angle=0 ,width=0.75\textwidth,keepaspectratio]{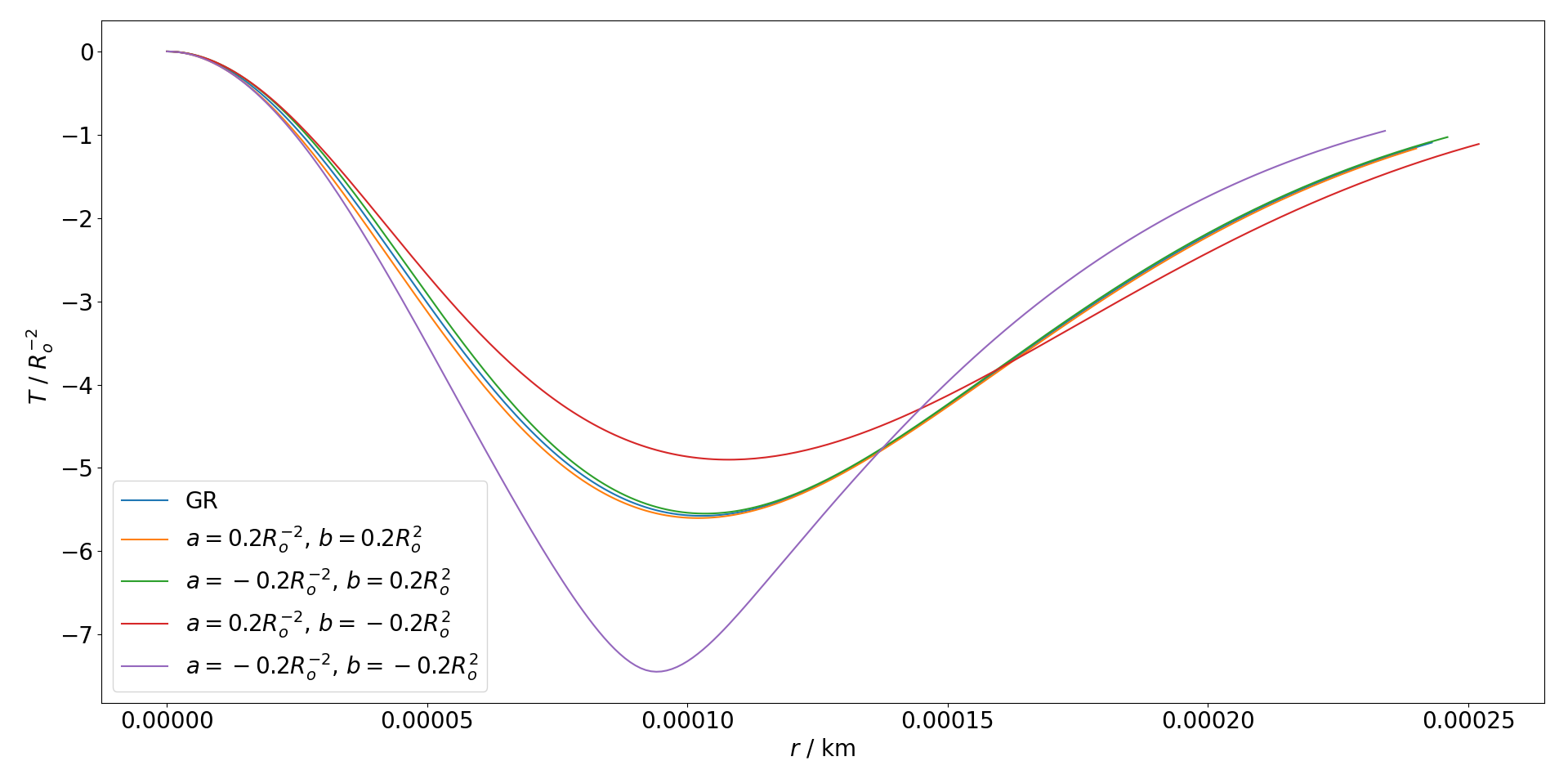}
\end{figure}

Moving to larger central densities, Fig.~\ref{tor18}, it is observed that the torsion scalar along stars in $f(T)$ gravity with a positive $b$ parameter tend towards the GR values. Meanwhile, the torsion scalar for a negative $b$ parameter gets weaker or stronger when compared to GR according to the signature of the parameter $a$, being positive or negative respectively. 

Regarding the viability of each class of mass-radius relationship shown in Fig.~\ref{compall} one also needs to look at the stability of the stellar configuration. In GR the stability of a branch in the mass-radius relationship may be determined from the Bardeen, Thorne, and Meltzer (BTM) criterion \cite{btm1,btm2,btm3}, which states that a stable mode becomes unstable if the mass-radius relationship has a stationary points which turns anticlockwise with increasing central density \cite{sta1}. Then, if we assume the BTM criterion to hold in $f(T)$ gravity, and that the stellar configurations are stable at low central densities, then only the mass-radius relationships with a positive $a$ and negative $b$ parameter exhibits no such turning points, such that all stellar configurations are stable. This, however, is problematic as the mass in this class of mass-radius relationships continues to grow unbounded with increasing central densities, implying that stellar collapse never occurs and so putting in question the viability of this class of parameter values. The other three types of mass-radius relationships all exhibit a turning point, marked by the maximum mass of the mass-radius relationship, from stable configurations to unstable ones.

\section{Data Set Description and Analysis}\label{sec:data}

To test the $f(T)$ modification we will use the following dataset \cite{dataset} consisting of white dwarfs in eclipsing binaries. As highlighted in \cite{wdpaper}, such systems enable the determination of a white dwarf's mass, radius and effective temperature without assuming a reference mass-radius relationship. This data set contains the mass, radius and effective temperature of 26 such white dwarfs, with their mass ranging between 0.3--0.8~M$_\odot$. The stars in the data set are divided into those with a helium core and those with a carbon/oxygen core. Moreover, Parsons et al. \cite{dataset} were able to constrain the envelop size for some of the stars into either a thin or thick one. For the remaining stars the analysis was ran twice, assuming a thin and a thick envelop.

\begin{figure}[H]
	\caption{A scatter plot of the data set, along with the Salpeter mass-radius relationships in GR. The temperature and envelop correction to Salpeter's EoS is also plotted as a shift in the data points, as described by Eq. \eqref{correction}.} 
	\centering
	\label{datacorr}
	\includegraphics[angle=0 ,width=0.75\textwidth,keepaspectratio]{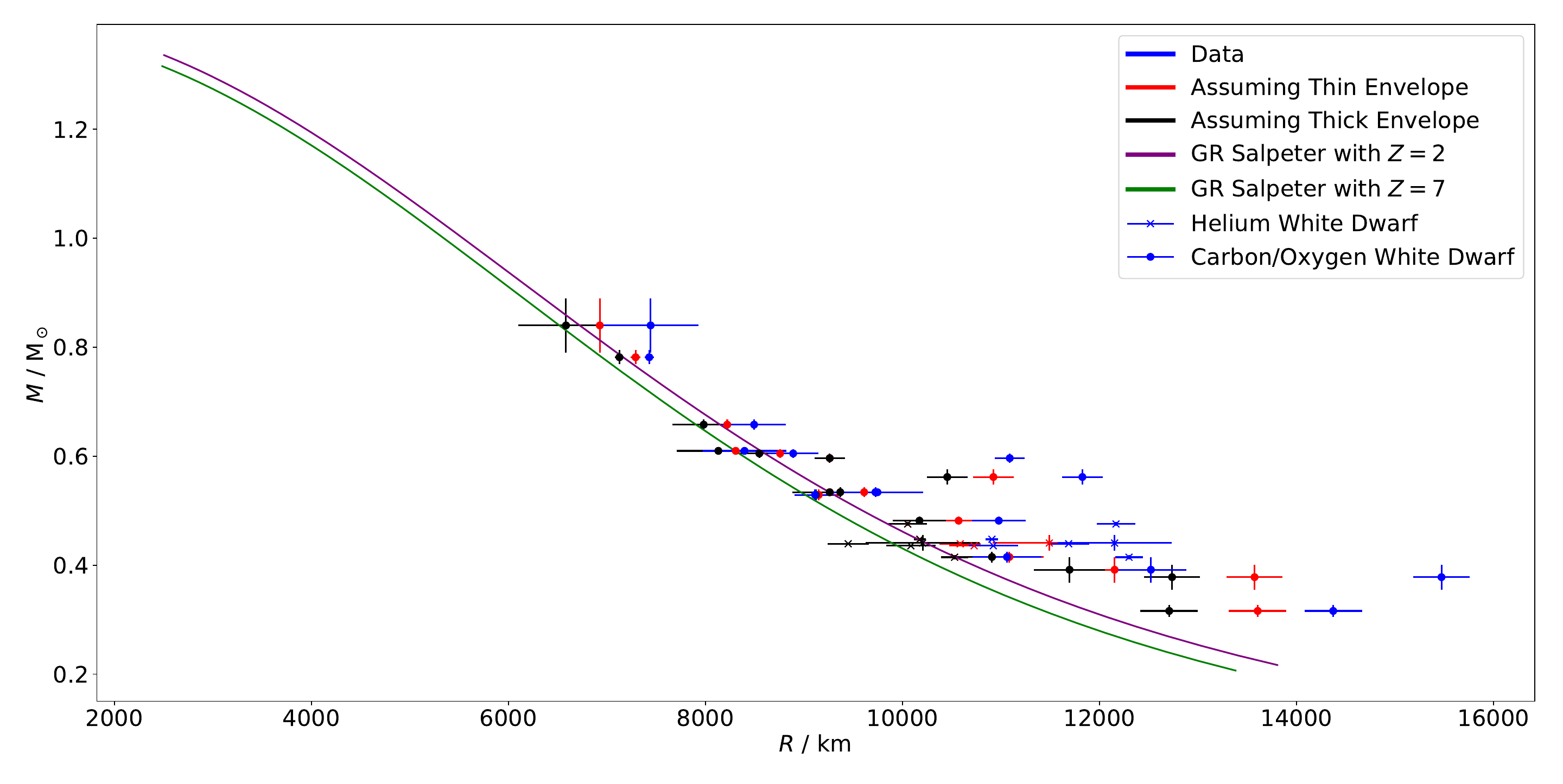}
\end{figure}

Fig.~\ref{datacorr} shows the data points along with the radius shift as the temperature and envelop correction to Salpeter's EoS. Shifting the radii of the data set, while equivalent to using a shifted mass-radius relationship for each star, allows us to better visualise the effects of the temperature and envelop correction as one mass-radius curve can then be used for all stars of the same composition. It should be noted that 5 data points had to be removed, as their temperature was outside the interpolation range provided by the evolutionary mass-radius relationship \cite{Panei:1999ji} used. The necessity of this correction can also be seen through this plot, where the actual data points, especially at low masses, deviate significantly from their respective zero-temperature mass-radius relationship. Moreover, here the effect that the envelop size has on the magnitude of the correction is clearly seen, with the shift being smaller when assuming a thin envelop. It should also be noted that when we assume a thin envelop for the white dwarfs which envelop size could not be constrained, the shifted data points lie further away from their respective GR zero-temperature mass-radius relationship, as such we expect that the deviation from GR required to fit the $f(T)$ to these points will be greater than when a thick envelop is assumed for these stars. In reality, we do not expect the envelop size of each of these stars to be the same, therefore these two analyses can be used to provide bounds for the extreme cases, with the fit to the thin envelop stars giving the maximum deviation from GR which may be necessary to explain white dwarf data.

For our analysis, we fit the mass-radius relationship given by Salpeter's EoS for the $f(T)$ parameters $a$ and $b$ to the shifted data $(M_i,R_i)_{i=1,...,21}$, by running a Markov Chain Monte Carlo (MCMC) analysis on the following modified chi-squared type test taking into account both uncertainty in the mass, $\Delta M_i$, and the radius, $\Delta R_i$, as the likelihood function \cite{chi}
\begin{equation}
    \chi^{2}=-\frac12\sum_{\mathrm{He}}\Bigg( \frac{  \left(M_{i}-    M_{\mathrm{Salpeter, Z=2}  }(a,b,R_i)  \right)^2   }{{\Delta M_{i}}^2 + {\left[\frac{\dd{M}}{\dd{R}}\right]_{i}}^2 {\Delta R_i}^2   }\Bigg) -\frac12\sum_{\mathrm{C/O}}\Bigg( \frac{  \left(M_{i}-    M_{\mathrm{Salpeter, Z=7}  }(a,b,R_i)  \right)^2   }{{\Delta M_{i}}^2 + {\left[\frac{\dd{M}}{\dd{R}}\right]_{i}}^2 {\Delta R_i}^2   }\Bigg)  \, .
\end{equation}
The separate information associated with the Helium and Carbon/Oxygen parts of the data are considered independently here, and the data does not feature any correlations.

\section{Constraint Analysis}\label{sec:constraint_analysis}

\begin{figure}[H]
	\caption{The posteriors and first quantile confidence regions obtained from the MCMC analysis.} 
	\centering
	\label{results1}
	\includegraphics[angle=0 ,width=0.75\textwidth,keepaspectratio]{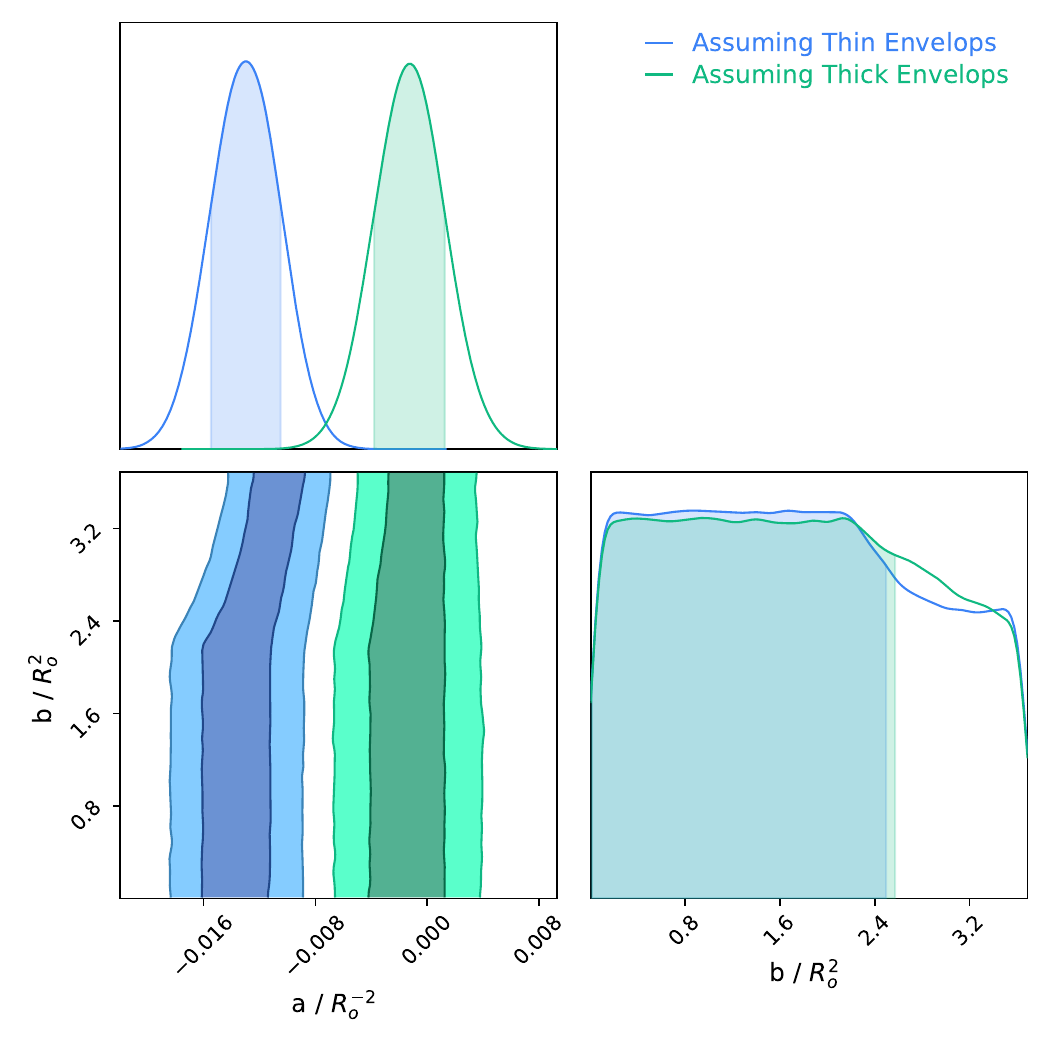}
\end{figure}

In this section, we present our results obtained from the MCMC analysis described in Sec.~\ref{sec:data}. Fig.~\ref{results1} gives the posteriors and confidence regions obtained from MCMC analysis, and the best fit values are given in Table~\ref{tab:model_params}. We first note that for both shifted datasets, corresponding to either an assumed thin or an assumed thick envelop for the stars whose envelop size could not be constrained, the shift required from GR is such that gravity is weaker. Another feature which is common to the posteriors for both shifted datasets is the flat posterior for $b$ between $0.2$ $R_\mathrm0^2$ and $2.6$ $R_\mathrm0^2$. Moreover, the fits are fairly Gaussian for the leading $a$ model parameter, while the constraints are less strong for the $b$ exponential parameter with a certain element of degeneracy with the $a$ model parameter.

\begin{table}[H]
    \centering
    \caption{Best Fit Parameter Values}
    \label{tab:model_params}
    \begin{tabular}{ccc}
        \hline
		Parameter & Assuming Thin Envelops & Assuming Thick Envelops  \\ 
		\hline
		$a$ / $R_o^{-2}$ & $\left(-12.9^{+2.4}_{-2.6} \right) \times 10^{-3}$ & $\left( -1.2^{+2.5}_{-2.6} \right) \times 10^{-3}$ \\ 
		$b$ / $R_o^{2}$ & $1.67^{+0.83}_{-1.66}$ & $0.0^{+2.6}_{-0.0}$ \\ 
		\hline
    \end{tabular}
\end{table}

\begin{figure}[H]
	\caption{The mass radius relationship of the median parameter value of $a= \left(-1.2^{+2.5}_{-2.6} \right)\times10^{-3}$ $R_\mathrm0^{-2}$ with upper and lower quantile values of $b$ required to fit the $f(T)$ model to shifted data set with an assumed thick envelop.} 
	\centering
	\label{thickmr}
	\includegraphics[angle=0 ,width=0.75\textwidth,keepaspectratio]{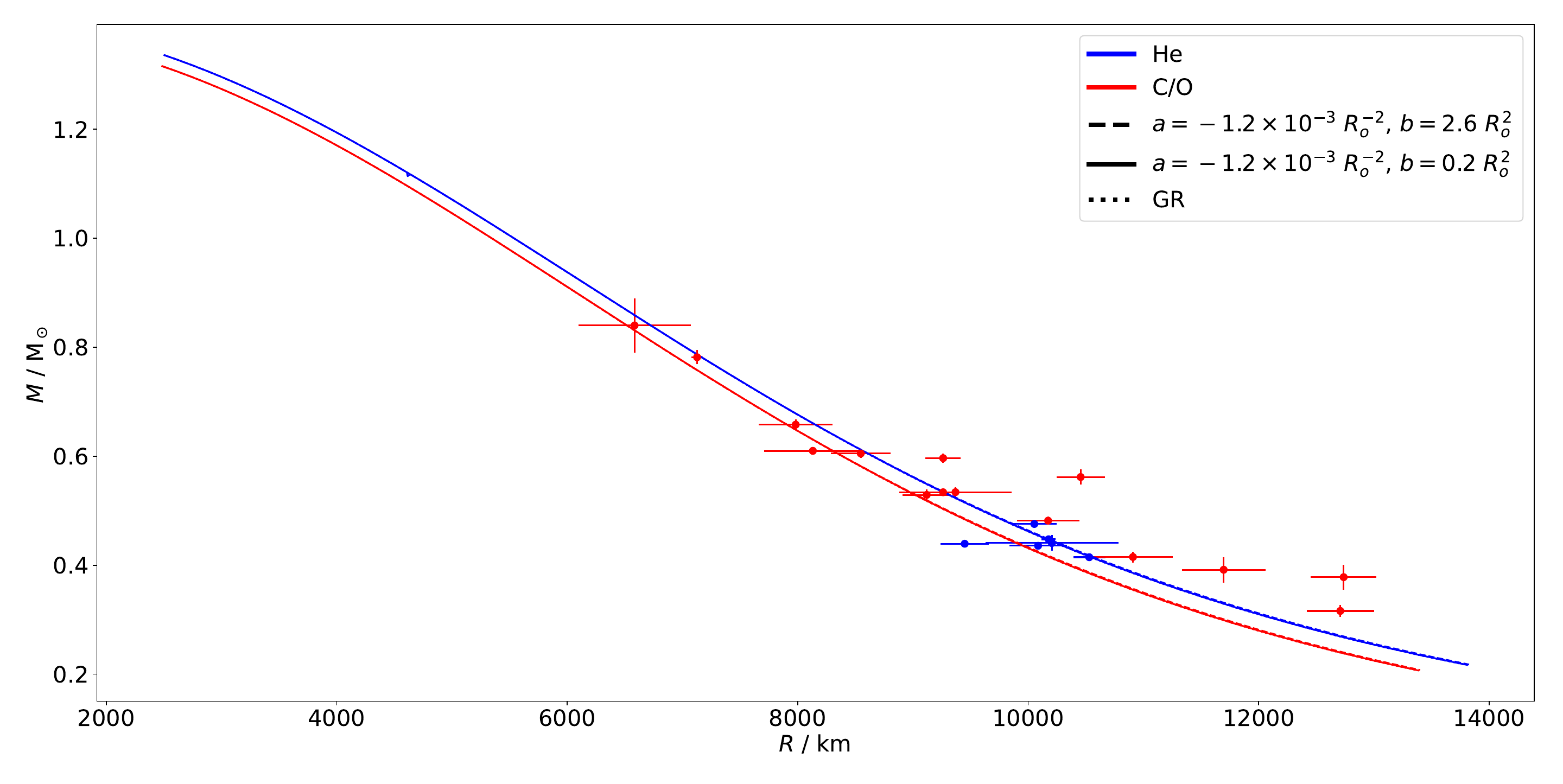}
\end{figure}

\begin{figure}[H]
	\caption{The mass radius relationship of the median parameter value of $a=\left(-12.9^{+2.4}_{-2.6} \right)\times10^{-3}$ $R_\mathrm0^{-2}$ with upper and lower quantile values of $b$ required to fit the $f(T)$ model to shifted data set with an assumed thin envelop.} 
	\centering
	\label{thinmr}
	\includegraphics[angle=0 ,width=0.75\textwidth,keepaspectratio]{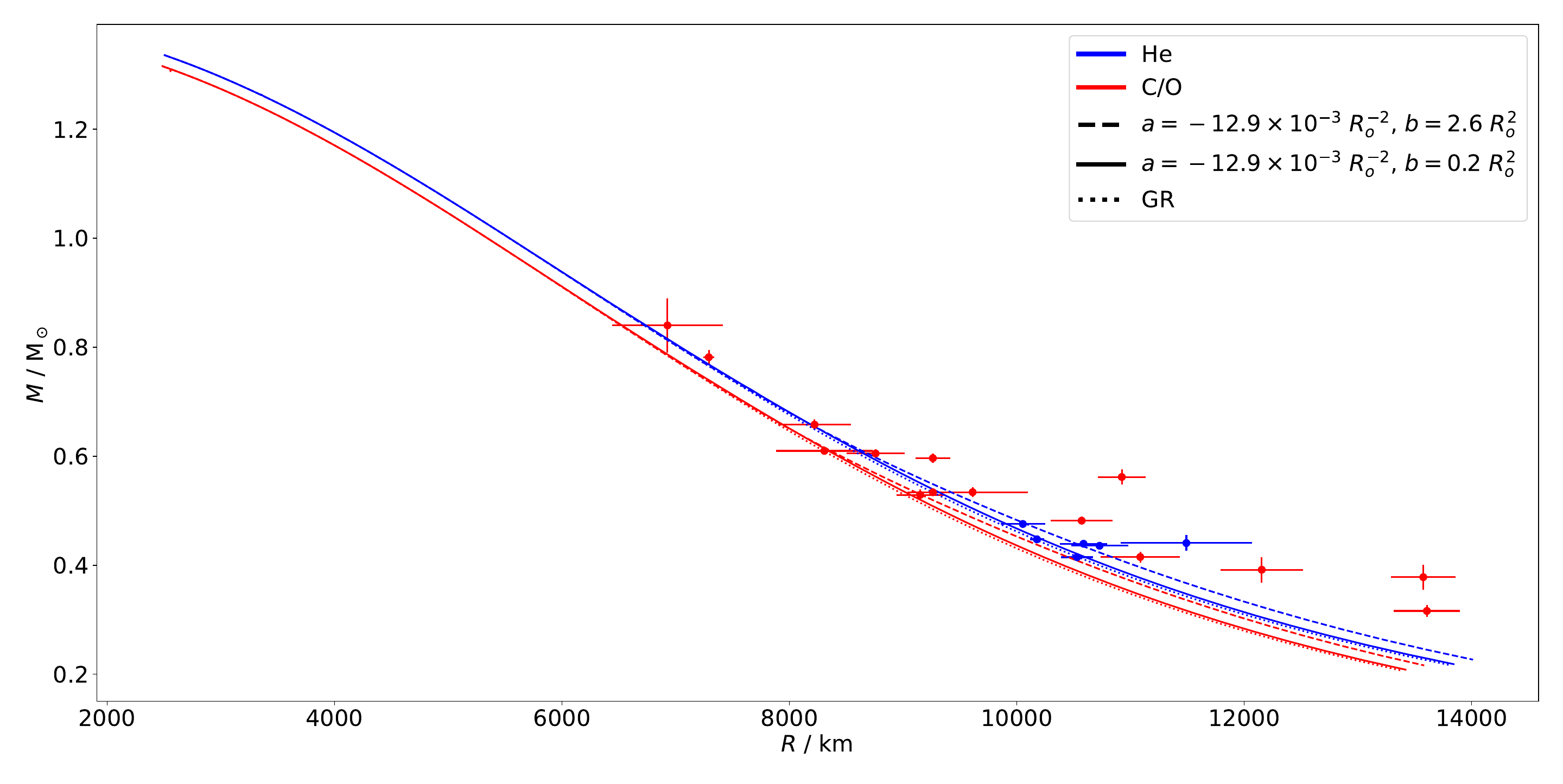}
\end{figure}

As shown in Figs.~\ref{thickmr} and \ref{thinmr}, it is evident that the mass-radius relationship is more sensitive to changes in $a$, with the largest deviations from GR occurring due to the larger value of $a$ required to fit the data to shifted data set with the stars having an assumed thin envelop.

From these figures, we also note the effect that our assumption on the envelop size has on the required modification from GR, with the biggest deviation required when the envelops are assumed to be thin. This is expected from Fig.~\ref{datacorr} and is in line with the results presented in Ref.~\cite{wdpaper}. As the actual envelop sizes will lie somewhere in between all thin and all thick we can use the upper quantile model obtained using the shifted data set with thin envelops to obtain an upper bound of $0<a<-15.5\times10^{-3}$ $R_\mathrm0^{-2}$ and $0<b<2.6 \, R_\mathrm0^{2}$ for the maximum deviation from GR expected from white dwarf data.

To compare our the best fit model to the cosmological best fit values presented in Ref~\cite{Briffa:2021nxg}, we convert all parameters in units of meters, such that the cosmological fits are of the order of $a\sim -\SI{e-31}{\metre^{-2}}$ and $b\sim \SI{e31}{\metre^{2}}$. The best fit value for $a$ assuming thin envelops is  $a= \left(-1.6 \substack{+0.31 \\ -0.32}\right)\times 10^{-23}$ m$^{-2}$ and $a= \left(-1.5 \substack{+3.1 \\ -3.2}\right)\times 10^{-24}$ m$^{-2}$ when thick envelops are assumed. Therefore, both envelop models give less strict bounds on $a$ than cosmological models. On the other hand the upper quantile value of $b$ for both envelop models is $b=-\SI{2e21}{\metre^{2}}$, which is closer to GR. Nonetheless, as the exponential model is more sensitive to $a$ rather than $b$, the mass-radius relationship corresponding to cosmological parameter values still lies between the GR and the best fit mass-radius relationships for both assumed envelop models, and thus within the bounds of the maximum deviation from GR as might be required to explain white dwarf data.

\section{Conclusion}\label{sec:conclu}

In this work, white dwarfs were explored through the prism of an exponential $f(T)$ gravity model model. This model has been shown to be interesting in the cosmological context of Hubble expansion \cite{Benetti:2020hxp,Xu:2018npu,Wang:2020zfv,Briffa:2021nxg,Briffa:2023ern}, as well as in the description of the large scale structure of the Universe \cite{Nesseris:2013jea,Anagnostopoulos:2019miu,Briffa:2023ozo}. In this work, we probe the strong field effects of this promising model, particularly in the context of white dwarfs. This was achieved by assuming a spherically symmetric background geometry~\eqref{met}, on which a perfect fluid is assumed for the stellar system, resulting in the system of equations of motion~(\ref{eom0}-\ref{eom2}) which describe the evolution of the stellar system for different boundary conditions. Combining the fluid conservation equation~\eqref{conserv} and boundary conditions in Eq.~\eqref{eq:boundary_cond}, the system can be fully realized using regular numerical methods.

At the numerical level, one already observes that for positive $b$, the exponential model decays for large central density values, while for negative values, divergences are possible. This is also born out by the data analysis which clearly shows a preference for positive values of $b$, as observed in the posterior plot shown in Fig.~\ref{results1}. While there are distinct values of the parameter $a$ depending whether a thin or thick envelop wall is considered, the values of $b$ are not sensitive to this assumption with almost overlapping constrain values showing consistency across both envelop models.

As future work, it would be interesting to consider more complex EoS structures, as well as the perturbative sector of spherically symmetric systems. It would also be interesting to explore how different $f(T)$ models are effected by the thin and thick envelop assumptions.

\acknowledgments{The authors would also like to acknowledge funding from ``The Malta Council for Science and Technology'' in project IPAS-2020-007. This paper is based upon work from COST Action CA21136 {\it Addressing observational tensions in cosmology with systematics and fundamental physics} (CosmoVerse) supported by COST (European Cooperation in Science and Technology). JLS would also like to acknowledge funding from ``The Malta Council for Science and Technology'' as part of the REP-2023-019 (CosmoLearn) Project.}

\bibliographystyle{unsrt}
\bibliography{references}

\appendix

\section{Appendix: Data Set}

In Table~\ref{data} we present the data set obtained from Ref.~\cite{dataset}, along with the shifted radii corresponding to both an assumed thin or thick envelop size where such an assumption was needed, where the shifted radius is referred to as $R_{\text{thin}}$ and $R_{\text{thick}}$ respectively. It is noted that for most stars the envelop size could not be constrained by Parsons et al., as the data was not precise enough to be more consistent with one or another envelop model.

\begin{table}[H]
\centering
\caption{Data Set Ref.~\cite{dataset} used along with temperature and envelop corrected radii following \cite{wdpaper}.}
\resizebox{\textwidth}{!}{%
\begin{tabular}{c|c|c|c|c|c|c|c|c|c}
Name             & $\theta$ / K & $M$ / M$_\odot$ & $M_{\text{Error}}$ / M$_\odot$ & $R_{\text{Data}}$ / km & $R_{\text{Error}}$ / R$_\odot$ & Composition & $R_{\text{thick}}$ / km & $R_{\text{thin}}$ / km & Envelop Size  \\ \hline
SDSS J0138-0016  & 3570  & 0.529         & 0.01                & 9119.75          & 0.0003                   & C/O         & 9119.75                 & 9150.6                 & Unconstrained \\
SDSS J1210+3347 & 6000  & 0.415         & 0.01                & 10909.1          & 0.0005                   & C/O         & 10909.1                 & 11085.9                & Unconstrained \\
SDSS J0024+1745  & 8272  & 0.534         & 0.009               & 9369.16          & 0.0007                   & C/O         & 9369.16                 & 9611.57                & Unconstrained \\
SDSS J1307+2156 & 8500  & 0.6098        & 0.0031              & 8131.52          & 0.00061                  & C/O         & 8131.52                 & 8307.86                & Unconstrained \\
SDSS J1123-1155 & 10210 & 0.605         & 0.0079              & 8547.35          & 0.00037                  & C/O         & 8547.35                 & 8758.39                & Unconstrained \\
CSS 41177B       & 11864 & 0.316         & 0.011               & 12708.5          & 0.00042                  & C/O         & 12708.5                 & 13606.5                & Unconstrained \\
SDSS J1329+1230  & 12491 & 0.3916        & 0.0234              & 11696.4          & 0.00052                  & C/O         & 11696.4                 & 12154.2                & Unconstrained \\
QS Vir           & 14220 & 0.7816        & 0.013               & 7127.15          & 0.00007                  & C/O         & 7127.15                 & 7291.36                & Unconstrained \\
CSS 40190        & 14901 & 0.4817        & 0.0077              & 10172.9          & 0.00039                  & C/O         & 10172.9                 & 10570.5                & Unconstrained \\
CSS 21357        & 15909 & 0.6579        & 0.0097              & 7982.69          & 0.00046                  & C/O         & 7982.69                 & 8221.11                & Unconstrained \\
V471 Tau         & 34500 & 0.84          & 0.05                & 6584.72          & 0.0007                   & C/O         & 6584.72                 & 6927.76                & Unconstrained \\
SDSS J0314+0206 & 46783 & 0.5964        & 0.0088              & 11089.458        & 0.00022                  & C/O         & 9262.14                 & 9262.14                & Unconstrained \\
GK Vir           & 50000 & 0.5618        & 0.0142              & 10455.1          & 0.0003                   & C/O         & 10455.1                 & 10923.5                & Unconstrained \\
SDSS J1021+1744 & 10644 & 0.5338        & 0.0038              & 9746.757         & 0.00032                  & C/O         & 9261.75                 & -                      & Thick         \\
CSS 41177A       & 22497 & 0.378         & 0.023               & 15472.368        & 0.00041                  & C/O         & 12737.6                 & -                      & Thick         \\
RRCae            & 7540  & 0.4475        & 0.0015              & 0.01568          & 0.00009                  & He          & 10177                   & -                      & Thick         \\
SDSS J1028+0931  & 12221 & 0.4146        & 0.0036              & 0.01768          & 0.0002                   & He          & 10529.9                 & -                      & Thick         \\
CSS 080502       & 17838 & 0.4756        & 0.0036              & 0.01749          & 0.00028                  & He          & 10053.4                 & -                      & Thick         \\
WD 1333+005      & 7740  & 0.4356        & 0.0016              & 0.0157           & 0.00036                  & He          & 10085.8                 & 10727.9                & Unconstrained \\
SDSS J0106-0014  & 13957 & 0.4406        & 0.0144              & 0.01747          & 0.00083                  & He          & 10205.3                 & 11491.3                & Unconstrained \\
SDSS J1212-0123 & 17707 & 0.4393        & 0.0022              & 0.0168           & 0.0003                   & He          & 9449.28                 & 10586.7                & Unconstrained
\end{tabular}%
}
\label{data}
\end{table}
\end{document}